\documentclass{article}
\usepackage{spconf,amsmath,graphicx, multirow}  
\usepackage{caption}
\usepackage{subcaption}
\usepackage{booktabs}
\usepackage{url}
\usepackage{algorithm}
\usepackage{algorithmic}
\usepackage{array}
\newcolumntype{P}[1]{>{\centering\arraybackslash}p{#1}}

\title{MELONS: generating melody WITH LONG-TERM STRUCTURE USING transformers and structure graph}
\name{Yi Zou$^{\ddagger \| }\sthanks{Equal contribution. This work was done when the two authors were on an internship at Kuaishou.} $ \quad Pei Zou$^{\mathsection\|\ast}$ \quad Yi Zhao$^{\|}$\sthanks{Corresponding author. zhaoyi07@kuaishou.com} \quad Kaixiang Zhang$^{\|}$ \quad Ran Zhang$^{\|}$ 
\quad Xiaorui Wang$^{\|}$ }

\address{$^{\ddagger}$  School of Information and Communication Engineering, Communication University of China\\
      $^{\mathsection}$ Department of Computer Science and Technology, Peking University\\
       $^{\|}$Kuaishou Technology Co., Beijing}

\begin{document}
\ninept
\maketitle
\begin{abstract}
The creation of long melody sequences requires effective expression of coherent musical structure. 
However, there is no clear representation of musical structure.
Recent works on music generation have suggested various approaches to deal with the structural information of music, but generating a full-song melody with clear long-term structure remains a challenge.
In this paper, we propose MELONS, a melody generation framework based on a graph representation of music structure which consists of eight types of bar-level relations.
MELONS adopts a multi-step generation method with transformer-based networks by factoring melody generation into two sub-problems: structure generation and structure conditional melody generation. Experimental results show that MELONS can 
produce structured melodies with high quality and rich contents.


\end{abstract}
\begin{keywords}
Music structure, melody generation, structure graph, transformer
\end{keywords}

\section{Introduction}
\label{sec:intro}
Music composition requires not only inspiration, but also obeying the rules of music theory. The restriction of musical structure is one of the most important rules for complete melody creation~\cite{langston1989six}. 
The structure of a complete melody piece is related to the musical form which affects the melody presentation along the time axis, for example, the repetition of motives, the transition between phrases, and the development between sections~\cite{zhang2021structure}.
In melody composition, the restriction of the structure makes the melodic line cohesive and comprehensible.
While aesthetically appealing short melody pieces can be generated using the recent deep learning based technologies~\cite{colombo2017deep, wu2019hierarchical}, creating long melodies with reasonable structure remains a challenge~\cite{briot2020deep, ji2020comprehensive}, which forms the motivation of this work. 


One of the obstacles in generating structured long music is that there is no clear approach to describe the structure technically. 
Some works regard the structure of music as simple repetitive patterns~\cite{waite2016generating, medeot2018structurenet}. 
However, the structures are primarily designed to reflect the dependencies among note sequences in different time-scales. 
Such dependencies include not only repetitions but also flexible and versatile developments.
Some researchers consider structure as a potential component of music and try to use various neural networks to discover it from music data automatically~\cite{roberts2018hierarchical, huang2018music, zhang2021melody}, but the generated music samples cannot show the reasonable and clear organization of structure that exists in human created music.
Many works use conditional information such as chord progressions~\cite{chen2019effect,guo2021hierarchical,zhang2021structure} and section tags~\cite{zhang2021structure, dai2021controllable} to control the structure of the generated music, but this requires parallel annotated data, which is not easy to collect.

There has been a long history of dividing the effort of generating structured music into different steps. Todd~\cite{todd1989connectionist} proposed to use two cascaded networks to construct new pieces of music with hierarchical structure representations. The networks learn both certain aspects of musical structure as well as obscure structural aspects from musical examples. Later, Hornel~\cite{hornel1998learning} proposed MELONET system, in which the first network learns to recognize musical structures while the second network predicts the notes.
PopMNet~\cite{wu2020popmnet} also employs two networks to generate structured pop melodies: a structure generation network used to generate the structural representations, and a melody generation net for generating music sequences based on the structural representation and chord progression. Different from~\cite{todd1989connectionist} and ~\cite{hornel1998learning}, 
PopMNet proposes to represent melody structure with graph, which consist of two typical relations known as repetition and sequence between pairs of bars. It utilizes adjacency matrices of the graph to model the structural variations of a long melody. 
Although PopMNet integrates melody structure into the generation process, the quality of the generated demos is quite limited. This is mainly due to the incomplete enumeration of the relations between bars. Repetition and sequence are just two categories among all common relations.  Further, the generation networks in PopMNet are based on CNN and RNN, which greatly limits the model's ability to generate time-related long sequences. 

In this paper, we propose a framework named MELONS which can generate full-song melody with long-term structure. Our contributions can be summarized as follows.
Firstly, instead of considering only simple repetitions and rhythm sequence, we construct a graph representation of melody structure with eight types of bar-level relations that appear frequently in the pop songs. These relations have covered most of the permissive music creation techniques including cadence and breath. Statistical analysis on multiple datasets from different sources has proved the versatility and rationality of our proposed relation types. 
In addition, we propose a multi-step generation method with transformer-based networks by factoring melody generation into structure generation and structure conditional melody generation.
Unlike PopMNet which is limited by a fixed 32-bar length and other works which require annotated structural information, our method can generate full-song melodies with flexible length, and is also annotation free.
Subjective evaluations show that our approach can generate harmonious melodies with reasonable structures.

\section{Proposed Approach}
\label{sec:pro}

\begin{figure*}[ht]
  \includegraphics[trim=1cm 20.4cm 1cm 0cm,clip,width=\textwidth]{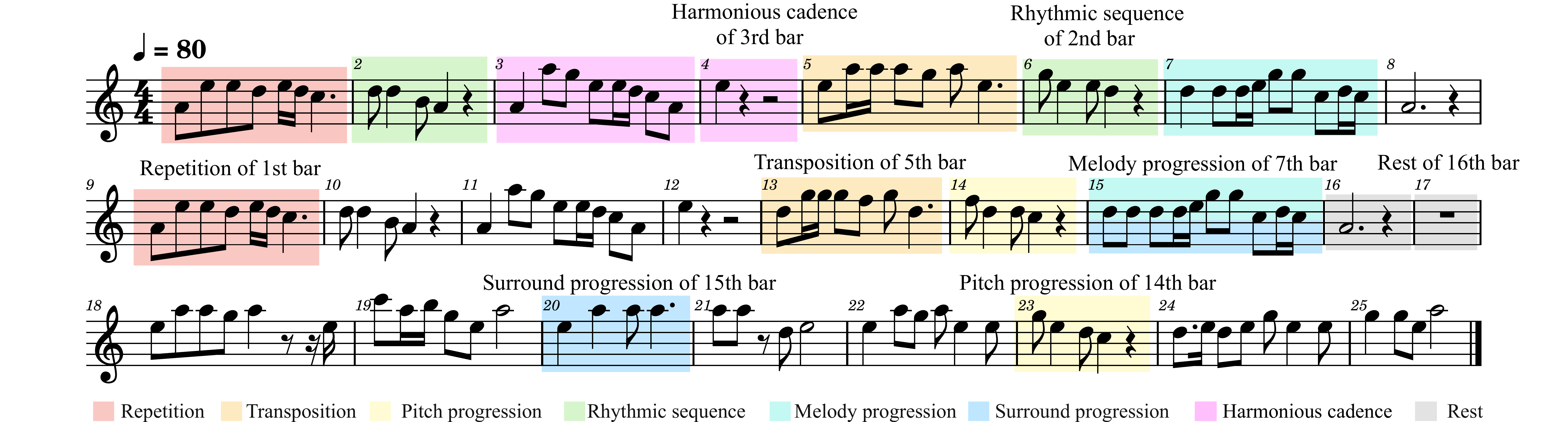}
\vspace{-0.3cm}
  \caption{Examples of the relations in music. The piece of melody comes from a Chinese song named ``Who are you dating tonight." Note that it shows only one example for each type of relation. }\label{img:score}
\vspace{-0.2cm}
\end{figure*}

\begin{table*}[htb]
    \centering
    \caption{Description of relation types. The priorities are sorted according to the similarity between the two bars in a relation}\label{tab:Description}
    \vspace{-0.2cm}
    \begin{tabular}{ccp{12.5cm}}
        \toprule
        Priority & Relation types & Description \\
        \midrule
        1 & Repetition & The current bar is the same as a previous bar.\\ 
        2 & Transposition & The current bar is the tonal transposition of a previous bar. \\
        3 & Pitch progression & Same rhythm. The similarity of pitch sequences between two bars is not less than 50\% \\ 
        4 & Rhythmic sequence & Same rhythm. The similarity of pitch sequences between two bars is less than 50\% \\
        5 & Melody progression & Two bars who have at least 3 consecutive notes of the same pitch and rhythm.\\
        6 & Surround progression & Surrounding notes(repeat more than 3 times in a bar) rise or fall by at least five semitones between two bars.\\
        7 & Harmonious cadence & A certain note in the current bar belongs to the local minimum point of the note density curve$^1$, and the pitch of this note belongs to the tonic chord of music key.\\
        8 & Rest & No notes in the current bar. \\
        \bottomrule
    \end{tabular}
    \vspace{-0.5cm}
\end{table*}

In this section, We firstly show a brief study on pairwise relations between bars, and then 
we introduce our proposed framework, MELONS, whose goal is to produce full-song melodies with long-term structures given melody motifs.

\subsection{Proposed bar-level relations}
\label{subsec:stru}

We employ the relations between pairwise bars to represent the musical structure. We take four beats as one bar and each beat equals to a quarter note.
Based on the knowledge of musical composition theory, we classify the bar-level relations  into three categories: repetition, development~\cite{kachulis2003songwriter} and cadence~\cite{kachulis2004songwriter}. 
After statistical analysis on more than 2500 pieces of pop music,  we further decompose these three categories into eight frequently appeared types of structural relations, which are defined in Table~\ref{tab:Description}. Fig.~\ref{img:score} shows an example of the relations in music score.

These relations have covered most common skills of pop music composition. They not only include complicated development methods, but can also reflect the progress of emotions (surround progression) and express the harmonious ending or breath of phrases (harmonious cadence). Moreover, these relations can be expressed by mathematical formulas\footnote{For more details, please refer to \url{https://yiathena.github.io/MELONS/}}, which facilitates their auto detection.

\begin{figure}[t]
  \centering
  \includegraphics[scale=0.22]{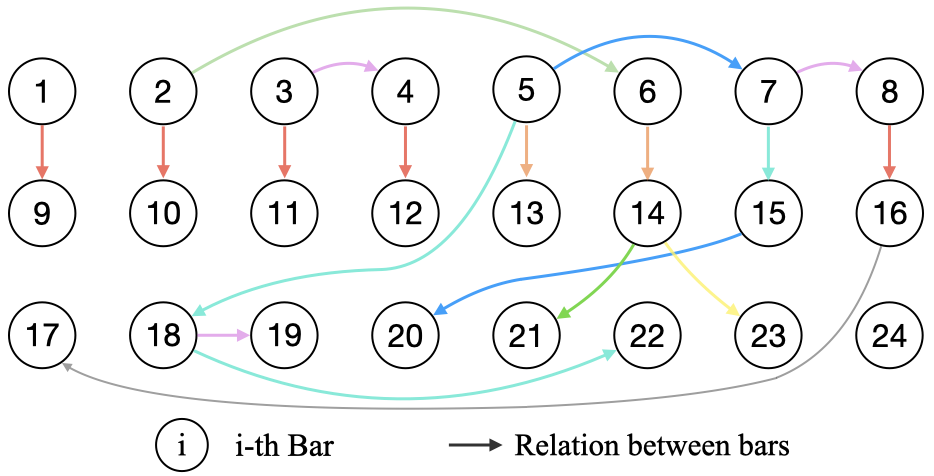}
   \vspace{-0.2cm}
  \caption{The structure graph of the melody in Fig.~\ref{img:score}. The color of the line represents the type of the relation. Note that it remains all of the relations finally kept by the structure graph. }
  \label{img:dag}
  \vspace{-0.5cm}
\end{figure}

We use a directed graph with multiple types of edges, which we call a structure graph, to describe the melody structure of a song. Bars and the relations between bars are represented as nodes and edges respectively, while types of the edges stand for types of the relations.
As is shown in Table~\ref{tab:Description}, the priority of relations is determined by the similarity between two bars.
To avoid redundancies in the structure graph, each bar is matched with its previous bars from near to far to find out the relation of the highest priority.
Fig.~\ref{img:dag} illustrates the structure graph of the melody provided in Fig.~\ref{img:score}. 


\subsection{Framework of MELONS}
\label{subsec:model}

\begin{figure}[t]
\begin{subfigure}[b]{.48\textwidth}
  \centering
  \includegraphics[width=\linewidth]{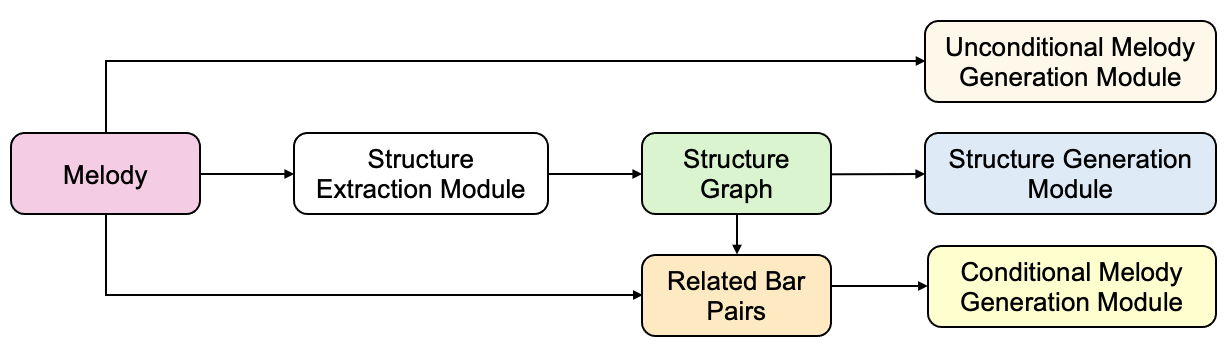}
  \caption{Framework of training}
\end{subfigure}
\begin{subfigure}[b]{.48\textwidth}
  \includegraphics[width=\linewidth]{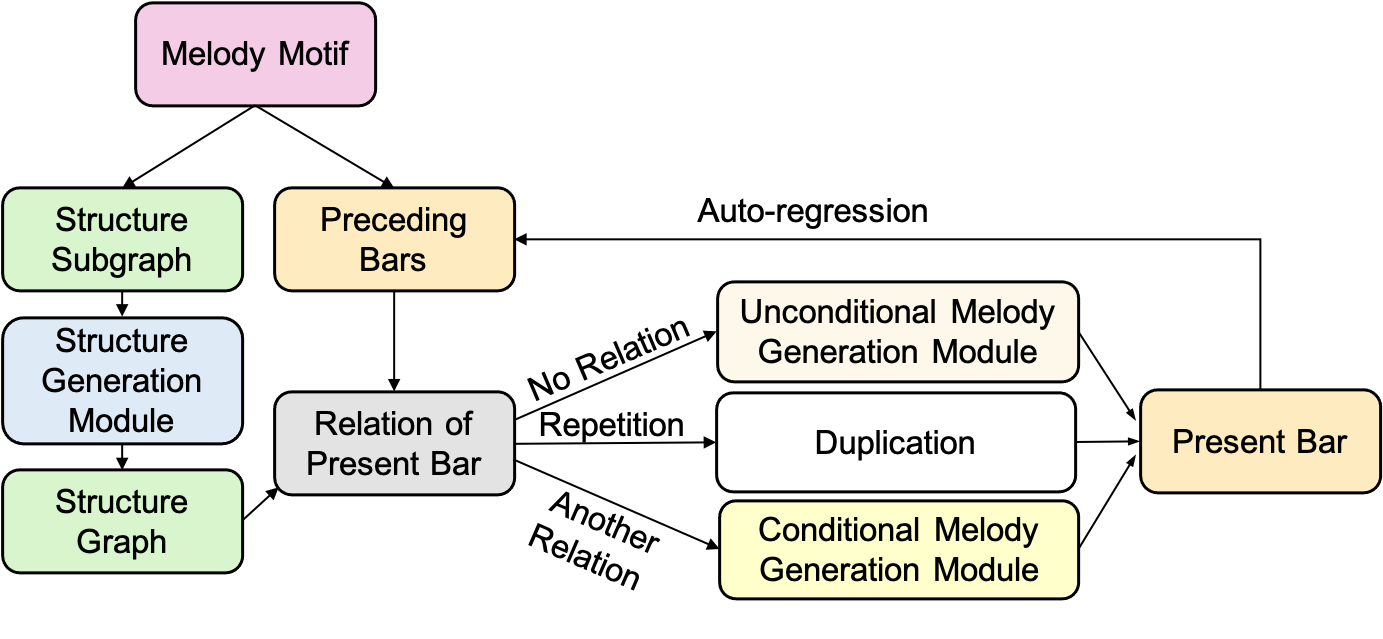}
  \caption{Framework of generation}
\end{subfigure}
 \caption{Proposed framework of MELONS. }\label{img:fra}
 \vspace{-0.5cm}
\end{figure}

The training and generation procedures of MELONS are shown in Fig.~\ref{img:fra}. We use the structure graph to bridge the gap between local pairwise relations and global music structure, and propose a multi-step melody generation framework to compose a full-song melody based on an 8-bar motif (which is the common length of a phrase). 

\textbf{Structure generation.} We extract structure graphs from the original songs to construct the dataset for structure generation. Inspired by GraphRNN~\cite{you2018graphrnn}, we model the structure graph as a sequence of relations, and generate the relations of a song using an auto-regressive transformer model. We use a token triple like $(i, j, t)$ to represent the edge(relation) from a previous bar $j$ to the current bar $i$ with a relation type $t$. 
By arranging all edges of a structure graph in ascending order of the index of the current bar, we get a list of token triples like $\{(i_1,j_1,t_1), (i_2,j_2,t_2), \cdots, (i_n,j_n.t_n)\}$, which is a sequence form of the structure graph. We adopt the tokenize method proposed in~\cite{hsiao2021compound} and predict the three tokens of an edge at one time step. As MELONS deals with full-song melody generation, we append an EOS relation at the end of each structure graph to mark the end of the song. At the inference time, edges in the structure graph would be generated autoregressively until the EOS is predicted, so the generated structure is not limited to a fixed length. Moreover, the subsequent relations are predicted conditioned on the music structure generated so far, leading to better cohesiveness. 

\textbf{Melody generation.} The melody generation part consists of an unconditional music generation module and a conditional generation module. We adopt the event-based music token representation applied broadly for music generation~\cite{huang2018music,huang2020pop,hsiao2021compound,zhang2021structure} in both two modules. The unconditional module is an auto-regressive Transformer model~\cite{hsiao2021compound} trained on the original melodies. The conditional generation module adopts a similar architecture, but training data is reorganized to build a conditional scenario, like CTRL~\cite{keskar2019ctrl} for conditional text generation and MuseNet~\cite{payne2019musenet} for music. For each relation $(i, j, t)$ in a structure graph, we treat the melody of the $i^{th}$ bar as the target and train the model to generate the $i^{th}$ bar conditional on the related $j^{th}$ bar and the relation type $t$. 
The final sequence has the form of \{[start of the melody context], (content of the melody context),[start of the related bar], (content of the related bar), [relation type], [start of the target bar], (content of the target bar)\}. The melody context here is formed by the 8 bars prior to the $i^{th}$ bar. Tokens before the [start of the target bar] token would be given at the inference time. The three modules mentioned above are combined together in the generation procedure. The detailed generation algorithm is illustrated in Algorithm 1.

\begin{algorithm} 
    \renewcommand{\algorithmicrequire}{\textbf{Input:}}
	\renewcommand{\algorithmicensure}{\textbf{Output:}}
	\caption{Generation procedure} 
	\label{alg3} 
	\begin{algorithmic}
	    \REQUIRE{melody motif $M = \{b_1, b_2, \cdots, b_8\}$}
	    \ENSURE{full-song melody $S = \{b_1, b_2, \cdots, b_n\}$}
	    \STATE $S \gets M$
	    \STATE Extract the structure of $M$ into relation list $R$
	    \REPEAT
	    \STATE Predict \emph{next relation} autoregressively and append \emph{next relation} to $R$
	    \UNTIL EOS relation predicted
	    \STATE $l \gets$ index of the last bar in $R$
	    \FOR{$i\leftarrow 9$ \TO $l$}{
	        \STATE Check relation $r = (i, j, t)$ of the $i^{th}$ bar in $R$
	        \IF {$r$ does not exist}
	        \STATE Generate the $i^{th}$ bar $b_i$ autoregressively with the unconditional generation module
	        \ELSIF{relation type $t$ is repetition}
	        \STATE $b_i \gets b_j $
	        \ELSE
	        \STATE Get melody context $C = \{b_{i - 8}, b_{i - 7}, \cdots, b_{i - 1}\}$ from $S$
	        \STATE Generate $b_i$ with the conditional generation module conditioned on $C$, $b_j$ and the relation type
	        \ENDIF
	    }
	    \STATE Append $b_i$ to $S$
	    \ENDFOR
	    \RETURN $S$
	\end{algorithmic} 
	
\end{algorithm}
\section{Experimental configurations}
\label{sec:exp-con}

\subsection{Dataset}
\label{subsec:db}
The datasets used in this work contain POP909~\cite{wang2020pop909} and Wikifonia\footnote{\url{http://www.wikifonia.org}}. They are public datasets of pop Chinese and Western songs, respectively. We keep melodies in 4/4 time signature as our training data and select 901 musical scores from POP909 and 1600 scores from Wikifonia with different pop music styles. We use a 9:1 method to allocate the training and testing set.

\subsection{Implementation details}
\label{subsec:exp-gr}
In MELONS, we choose the linear Transformer~\cite{katharopoulos2020transformers} as the backbone architecture for sequence modeling, considering its linear complexity in attention calculation. The structure generation module consists of 4 self-attention layers with 4 heads and 256 hidden states. The inner layer size of the feed-forward part is set to 1024. The melody generation modules consist of 6 self-attention layers with 8 heads and 512 hidden states, with the inner layer size set to 2048. We adopt the \underline{c}om\underline{p}ound word (CP) method~\cite{hsiao2021compound} for sequence representation and tokenization. The structure generation module employs a token set including three types of tokens: index of the current bar, index of the related bar and relation type, and the embedding size is 128, 128 and 32 respectively. The unconditional melody generation module uses a token set $(type, beat, tempo, pitch, duration)$ to serialize melodies, with embedding size as $(16, 128, 64, 256, 128)$. We additionally use $track(size=16)$ to mark the different sections in the conditional generation module, and use $relation type(size=32)$ to impose the control. The embedding sizes are chosen based on the vocabulary sizes of different token types, and the embedded tokens describing the same object would be concatenated together and linearly projected to the same size of the corresponding module's hidden state. We use the Adam optimizer~\cite{kingma2014adam} with a learning rate of $10^{-4}$ for all these models. We apply dropout with a rate of 0.1 during training. 

\section{Experimental Evaluation}
\label{sec:exp}

\subsection{Objective evaluations}
\label{subsec:obj}

\begin{table}
\centering
\caption{Percentages of relation types in different datasets.} 
 \vspace{-0.2cm}
\begin{tabular}{c|P{0.98cm}P{0.98cm}P{0.98cm}P{0.98cm}}
\toprule
Relation types  & POP909 & Wikifonia & Collected & MELONS \\
\midrule
Repetition & 44\%   & 47\%     & 50\%   & 50\%    \\
Transposition & 1\%     & 3\%     & 2\%   & 1\%    \\
Pitch progression & 3\%     & 4\%     & 3\%   & 4\%    \\
Rhythmic sequence & 6\%     & 8\%     & 6\%   & 11\%    \\
Melody progression & 11\%     & 3\%     & 6\%   & 6\%    \\
Surround progression & 5\%           & 2\%     & 3\%   & 2\%    \\
Harmonious cadence & 6\%           & 5\%     & 5\%   & 5\%    \\
Rest & 6\%           & 8\%     & 7\%   & 4\%    \\
In total & 81\%           & 81\%     & 82\%   & 83\%    \\
\bottomrule
\end{tabular}
 \label{tab:objective}
 \vspace{-0.5cm}
\end{table}

In order to verify the versatility of the relationship we define in Table~\ref{tab:Description}, we additionally collect 1160 pop music in different styles and regions from the Internet, and also collect thousands of melodies generated by the MELONS. 
We automatically extract all relations of each dataset. 
Then we perform statistical analysis on relations in each dataset. The proportion is defined as the ratio of the number of bars with a certain relation to the total number of bars.
Table~\ref{tab:objective} shows the statistical results of the proportions. We can see that the distribution of the relations across the four dataset are roughly close (correlation coefficients $>0.98$ for all pairs). This demonstrates the universality and stability of our proposed structure representations.



We also want to see whether our melody generation network can  produce target bars with correct relations under specified conditions.
After comparing the structure graphs extracted from the generated melodies with the graphs used as conditions, we get the ratio of correctly generated bars of each relation. Among all the relations, rhythmic sequence owns the highest accuracy$(92.88\%)$ while transposition shows the lowest performance$(77.33\%)$.


%

\subsection{Subjective evaluations}
\label{subsec:sub}
We compare the performance of MELONS with another two systems including CP-Transformer (SOTA method for unconditional music generation without control on structure) and PopMNet with different configurations. 
The structure graph used as the condition for the melody generation network can either be extracted from the real music (marked as \textbf{R}) or generated by the structure generation net (no mark).
The relations in the structure graph can be either a simple repetition and rhythm sequence as PopMNet suggested  (marked as \textbf{B}) or the more sophisticated relation set as described in ~\ref{subsec:stru}  (no mark).
We also include human created melodies for reference.

For the listening test, we randomly select 10 motifs from the testing set and generate sets of melodies corresponding to the above experimental systems\footnote{Audio samples of generated melodies are available at \url{https://yiathena.github.io/MELONS/}}. We invite 12 listeners who are professionals in music to evaluate 80 samples. The subjects are asked to rate the melodies on a five-point scale varying from very bad to excellent on the following four metrics: 1) \textbf{Structure}: Does the melody have clear structure? 2) \textbf{Richness}: Does the melody have rich content? 3) \textbf{Pleasure}: Does the melody sound melodious? 
4) \textbf{Overall}: What is the overall quality of the melody?

\begin{table}
\caption{The average mean opinion scores on the four evaluation metrics over all experimental groups.}
 \vspace{-0.2cm}
\begin{tabular}{@{}c|cccc@{}}
\toprule
Model     & Structure & Richness & Pleasure & Overall \\ \midrule
CP-Transformer       & 2.41           & 2.28     & 2.37   & 2.13    \\
PopMNet & 2.31           & 2.46     & 2.18   & 2.17    \\
PopMNet-R & 2.54           & 2.30     & 2.27   & 2.29    \\
MELONS-B & 2.77           & 2.60     & 2.39   & 2.34    \\
MELONS-B-R & 2.81           & 2.71     & 2.41   & 2.43    \\
MELONS   & 3.53           & 3.29     & 3.20   & 3.20    \\
MELONS-R   & 3.60           & 3.37     & 3.26   & 3.22    \\
Human     & 4.54           & 4.34     & 4.34   & 4.39    \\ \bottomrule 
\end{tabular}
\label{tab:mos}
\end{table}

\begin{figure}[t]
  \centering
  \includegraphics[scale=0.6]{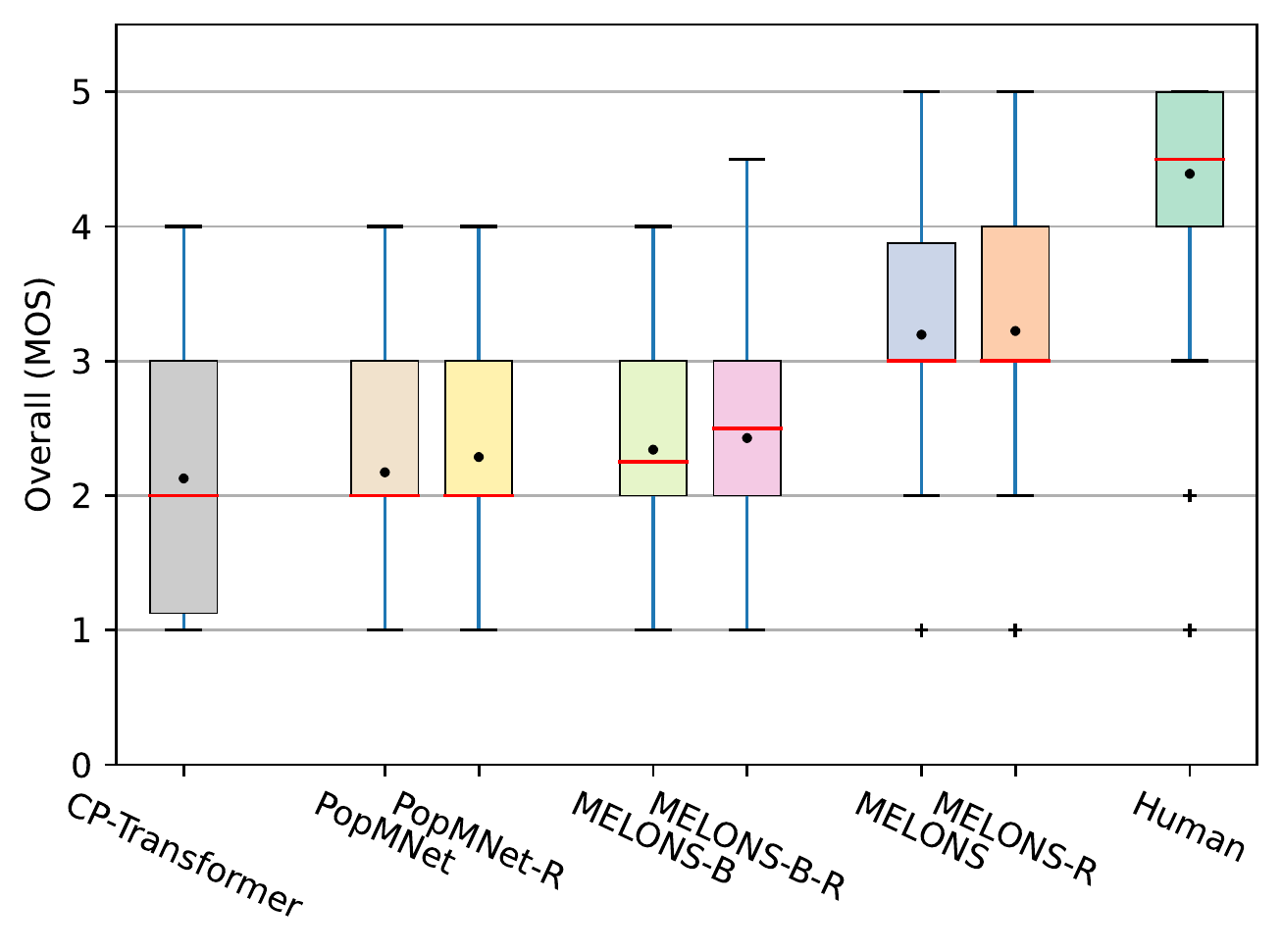}
  \vspace{-0.4cm}
  \caption{Box plots on MOS scores of overall. Black dots represent mean score.}
  \label{img:boxplot}
  \vspace{-0.5cm}
\end{figure}

Table~\ref{tab:mos} shows the average mean opinion scores on the four evaluation metrics. 
The distributions of the results of the four metrics are relatively similar, which to a certain extent illustrates the significance of musical structure in the overall evaluation of music.
Hence, we only show the box plots on the overall metric in Fig.~\ref{img:boxplot}.  T-tests with Holm-Bonferroni correction ($\alpha = 0.05$) were also conducted for all evaluation metrics.

The results demonstrate the effectiveness of our proposed approach. MELONS and MELONS-R outperform the other systems on all metrics by a large margin ($p<1e-6$ for all evaluation metrics). 
Among all the systems, CP-transformer archives the lowest evaluation. This indicates that, remarkably, melody generation can benefit from structural information. 
Furthermore, the performance of MELONS systems which use the sophisticated structure graph is obviously better than those using the basic structure graph proposed by PopMNet, which shows the superiority of our proposed structure representation.  
We also see that MELONS-B related systems are better than the PopMNet related systems, indicating that MELONS could generate better melodies when using the same structural information as prior knowledge. There is still an obvious gap between generated melodies and human created music, leaving large room for improvement. We notice that differences between models using the generated structure graph and those using the real structure graph are not significant.
This suggests that music structure generated by MELONS has close quality compared to artificially designed music forms. Thus our future work will focus more on improving the melody generation model.

\section{Conclusions}
\label{sec:con}
MELONS is a transformer-based framework which generates melodies with long-term structures by using a structure generation net and a melody generation net. 
One of the key ideas is to represent structure information using graph which consists of eight hierarchical relations among pairwise bars.
Evaluations show the effectiveness of such representations. 
The subjective evaluation results suggest MELONS can generate enjoyable melodies with obvious structure and rich contents. 

\newpage
\bibliographystyle{IEEE.bst}
\bibliography{mybib.bib}

\begin{figure*}[ht]
\centering
\includegraphics[width=\textwidth]{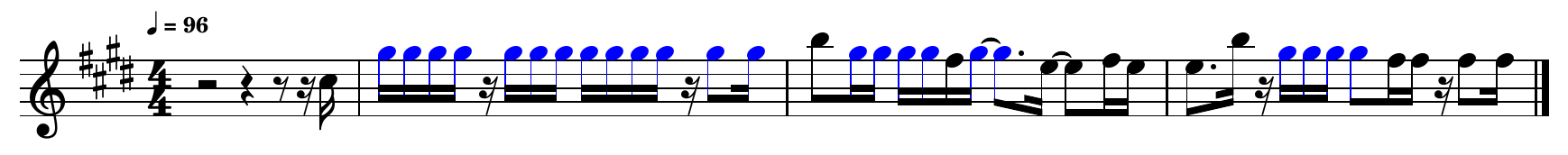}
\caption{Examples of surround notes (marked in blue in the figure).  The melodic line progresses and moves around the surround notes. The piece of melody comes from a pop song named ``Shape of you". }
\label{img:shapeofyou}
\end{figure*}

\newpage

\section{Appendix}
\label{sec:appendix}

This chapter is a supplement to Table~\ref{tab:Description}. Here we describe the definitions of some concepts that appear in the table. 

Since the relation types are determined by the rhythm and pitch sequence within bars, here we firstly explain the definitions of rhythm and pitch sequence used in this paper.
We use $R_i$ and $P_i$ to represent the rhythm  and pitch sequence of the $i^{th}$ bar, respectively. There are a set of  rhythm and pitch as follows:
\begin{equation}
R_i = \{ r_{i,1}, r_{i,2}, ... ,r_{i,k} , ... ,r_{i,n} \}
\\
P_i = \{ p_{i,1}, p_{i,2}, ... ,p_{i,k} , ... ,p_{i,n} \}
\end{equation}
where $k$ is the position of notes, and $r_{i,k}$ and $p_{i,k}$ represent the rhythm and pitch of the $k^{th}$ note within the $i_{th}$ bar, respectively. In order to show the rhythm in a mathematical way, we use the relative start time of each note in a bar to describe rhythm. In our experiment, the sixteenth note is used as the basic time unit.
$p_{i,k}$ is the midi index of the $k^{th}$ note in bar, for example, the index 60 represents the pitch C4 in midi protocol.
$n$ is the number of notes in a bar.
Taking Fig~\ref{img:score} as an example, the rhythm sequence and pitch sequence of the 14th bar and the 23rd bar are:
$R_{14} = \{ 0, 2, 6, 8 \}$, $P_{14} = \{ 77, 74, 74, 72 \}$, $R_{23} = \{ 0, 2, 6, 8 \}$, $P_{23} = \{ 79, 76, 74, 72 \}$.

\textbf{Pitch progression and rhythmic sequence.} The differences between pitch progression and rhythmic sequence are mainly existing in the similarity coefficient of the pitch sequences  in two bars.  
If there is a bar $i$ and another bar $j$, where $j>i$, and the number of notes in $j$  is $n$, the similarity of the bar $i$ and bar $j$ is determined by the number of the notes those have the same pitch:
\begin{align}
Similarity = \frac{\sum_{k=1}^n (p_{i,k} == p_{j,k})}{n}
\end{align}
If the total number of notes in $j-th$ bar is less than $n$, we set $p_{j,k}$ as 0.
If the similarity of pitch sequences between two bars is not less than 50\%, and their rhythm sequences are the same, we call the second bar as a pitch progression of the first bar.
For example, for the $14th$ bar and the $23rd$ bar in figure~\ref{img:score}, their rhythms are the same $R_{23}=R_{14}$ and their similarity is 0.5 since $n=4$ and $p_{23,3} = p_{14,3}, p_{23,4} = p_{14,4}$. Thus the $23rd$ bar is the pitch progression of the $14th$ bar because the similarity coefficient of the pitch sequence between the two is less than 50\%. 
In contrast, for the 6th bar and the 2nd bar in figure~\ref{img:score}, the similarity coefficient of the pitch sequence between the two is less than 50\% $(R_i=R_j, similarity=0)$, thus the 6th bar is the rhythmic sequence of the 2nd bar.



\textbf{Melody progression.} 
To find the bars that are written based on melody progression, we utilize an algorithm which is similar to finding the  longest common subsequence between two sequences. 
A common subsequence of two sequence is an ordered subsequence that is common to both sequences.
Assume $R_{i,j}$ is the longest common subsequence of rhythm  between bar $i$ and $j$,  and $P_{i,j}$ is the longest common subsequence of pitch sequence between bar $i$ and $j$, then $R_{i,j}$ and $P_{i,j}$ can be represented as follows:
\begin{equation}
\begin{split}
R_{i,j} = \{ r_{i,k} ,r_{i,k+1} , ... ,r_{i,m} \},  i>j, m\leq n\\
P_{i,j} = \{ p_{i,k} ,p_{i,k+1} , ... ,p_{i,m} \},  i>j, m\leq n
\end{split}
\end{equation}
if $R_{i,j}=P_{i,j}$ and $ m\ge 3 $ , the $i^th$ bar is the melody progression of  the $j^th$ bar.

\begin{figure}[t]
  \centering
  \includegraphics[scale=0.3]{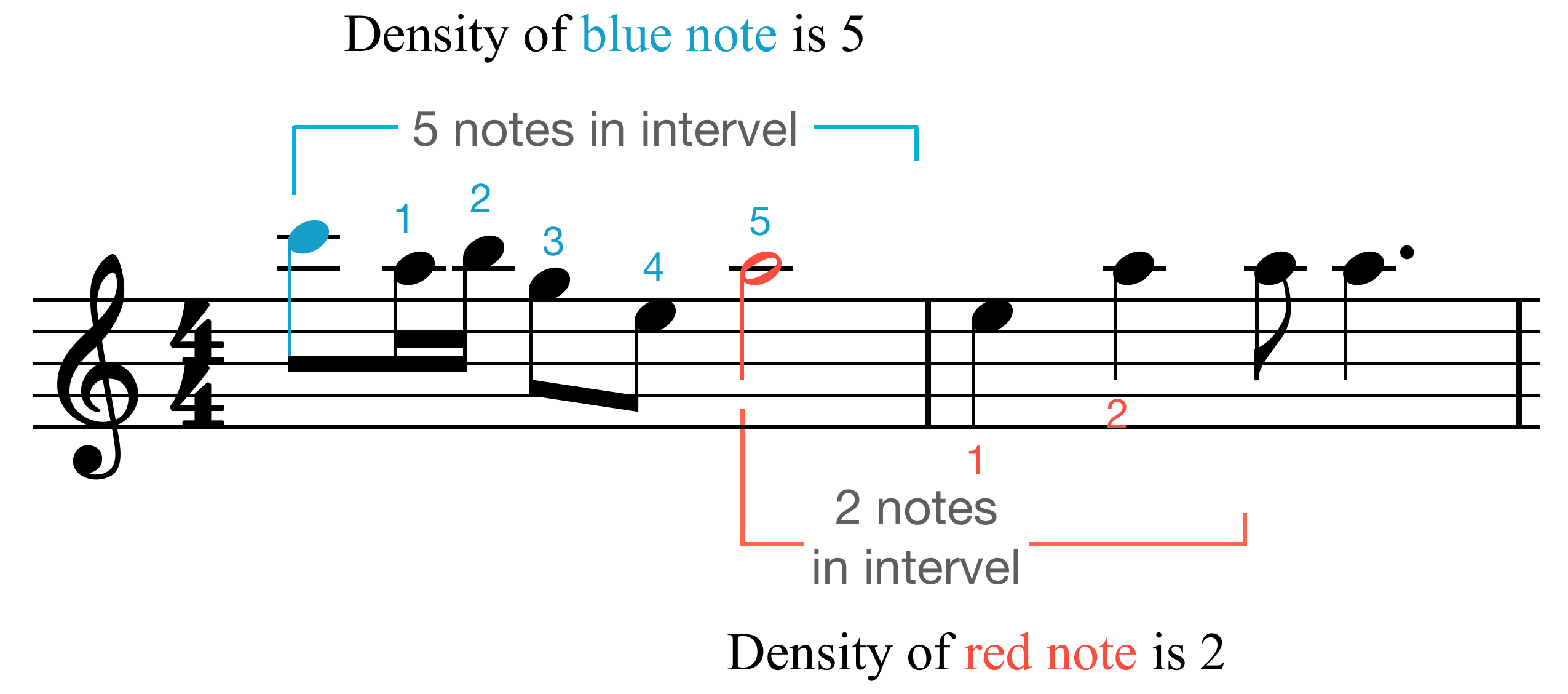}
  \caption{Examples of the note density in 19th bar, music notes on the interval boundary will not be counted as density. The red note is local minimum points of note density in this score }
  \label{img:d_note}
  \vspace{-0.3cm}
\end{figure}

\textbf{Harmonious Cadence.} Harmonious cadence is used to represent the harmonious end or breath of phrase. 
We employ note density curve to detect the position of harmonious cadence. The note density curve is formed by the position and density of the notes. Since the position of the notes is independent, the density of the notes is a function of the independent variable.
The density indicates the number of notes in the time interval that the note belongs to.
The lower bound of the time interval is the open interval of the notes' start time. The length of the interval is set as one bar.
Local minimum points of the note density curve is the local sparsest position in the melody, indicating the end of the phrase and breathing. 
Fig~\ref{img:d_note} shows an example of the music note density of the 19th bar.

\begin{figure}[t]
  \centering
  \includegraphics[scale=0.5]{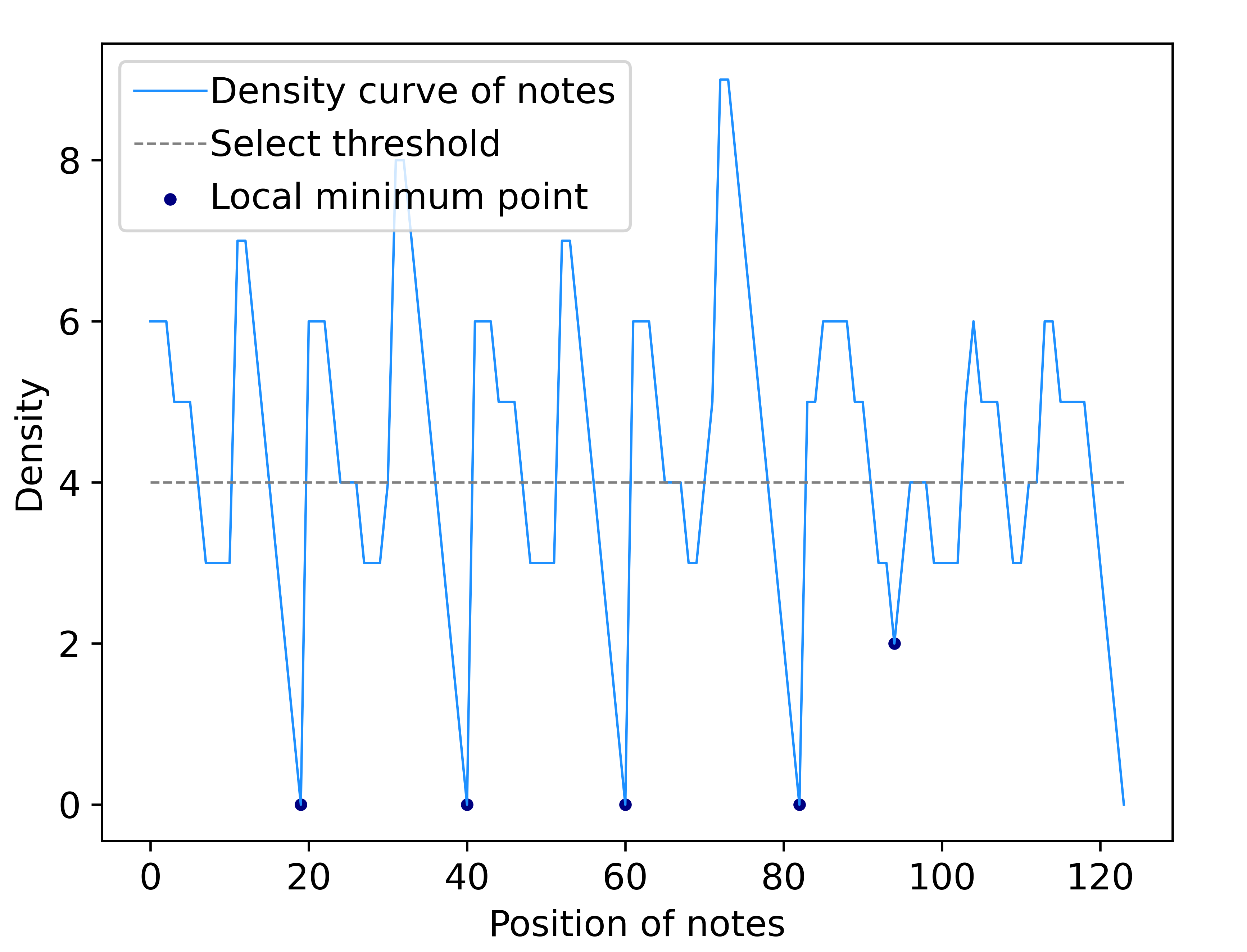}
  \caption{Note density curve for score in Fig~\ref{img:score}.}
  \label{img:density}
  \vspace{-0.3cm}
\end{figure}

The note density curve extracted from the melody of Fig.~\ref{img:score} is shown in Fig.~\ref{img:density}.
The fifth local minimum point in Fig.~\ref{img:density} is the red note shown in Fig.~\ref{img:d_note}, 

After analyzing the curve, we select the notes that belong to the tonic chord of music key from the local minimum points. 
This idea comes from a melody writing method called Cadence, which allows the phrase to create a sense of resolution. 

\textbf{Surround progression.} Surround progression is used to manifest the progression of emotions in music. 
In rap, rock and pop music, the melody in a phrase usually moves around a certain note, example is shown in Fig~\ref{img:shapeofyou}.

Raising the pitch sequence of the melody moving around a certain note is the easiest way to promote emotions. 
In this work, the notes that repeat more than 3 times in bar are called surround notes.
The method to find the type of Surround progression for the current bar is as follows:
First, eliminate the higher-priority relation types, and then determine whether there are surrounding sounds in the current bar. Finally, look for bars that also have surrounding sounds from near to far, and the surrounding sounds of this bar rise or fall at least five semitones.

\end{document}